\documentclass[prd,twocolumn,nofootinbib]{revtex4-1}
\usepackage{amsfonts}
\usepackage{amsmath}
\usepackage{amssymb}
\usepackage{bm}
\usepackage{dcolumn}
\usepackage[dvips]{graphicx}
\usepackage{subcaption}
\usepackage[normalem]{ulem}
\usepackage{latexsym}
\usepackage{rotating}
\usepackage[colorlinks=true]{hyperref}
\usepackage{xspace} 
\usepackage[usenames]{color}
\usepackage[dvipsnames]{xcolor}
\usepackage{mathrsfs}
\usepackage{multirow}
\usepackage{pifont}
\usepackage{enumitem}
\usepackage{color}
\usepackage{url}
\usepackage{array}
\usepackage{appendix}
\usepackage[utf8]{inputenc}
\usepackage{comment}
\usepackage{placeins} 
\definecolor {darkgreen}{rgb}{0.2,0.7,0.2}
\definecolor{purple}{rgb}{0.5,0,0.5}

\usepackage{graphicx}
\usepackage{epsfig}
\usepackage{slashed}
\usepackage{braket}
\usepackage{tensor}

\usepackage{cases}

\def\bec{\begin{center}}
\def\eec{\end{center}}
\def\beq{\begin{equation}}
\def\eeq{\end{equation}}
\def\bea{\begin{eqnarray}}
\def\eea{\end{eqnarray}}

\begin{document}

\title{Euclidean Dynamical Triangulations Revisited}


\author{Muhammad Asaduzzaman}
\email{masaduzz@syr.edu}
\affiliation{Department of Physics, Syracuse University, Syracuse, NY 13244, USA.}

\author{Simon Catterall}
\email{smcatter@syr.edu}
\affiliation{Department of Physics, Syracuse University, Syracuse, NY 13244, USA.}

\begin{abstract}
We conduct numerical
simulations of a model of four dimensional quantum gravity
in which the path integral over
continuum Euclidean metrics is approximated by a sum over
combinatorial triangulations.  At fixed volume the model
contains a discrete Einstein-Hilbert term with coupling $\kappa$ and
local measure term with coupling $\beta$ that weights triangulations
according to the number of simplices sharing each vertex.
We map out the phase diagram in this two dimensional parameter
space and compute a variety of observables that yield information on
the nature of any continuum limit. Our results are consistent with a line of
first order phase transitions with a latent heat that decreases as $\kappa\to\infty$.
We find a Hausdorff 
dimension along the critical line that
approaches $D_H=4$ for large $\kappa$ and a spectral dimension that is consistent with $D_s=\frac{3}{2}$ at short distances. These
results are broadly
in agreement with earlier works on Euclidean dynamical triangulation
models which utilize degenerate triangulations and/or different measure terms
and indicate that such models exhibit a degree of universality. 

\end{abstract}
\date{\today}
\maketitle

\newpage
\section{Introduction}
There are many proposals for quantizing four dimensional gravity - see
the reviews \cite{ashtekar2021short,loll2019quantum,weinberg1979ultraviolet,niedermaier2006asymptotic}. In this paper we explore one such approach known as Euclidean Dynamical Triangulation (EDT) in which the continuum path integral is replaced by a discrete sum over
simplicial manifolds. This approach is similar in spirit
to the Causal Dynamical Triangulation (CDT) program \cite{loll2019quantum,ambjorn2013causal} after relaxing the
constraint that each triangulation admit a discrete time
slicing. In practice we restrict to triangulations
with equal edge lengths and fixed topology. In addition we only
include so-called combinatorial triangulations in the discrete path integral
which guarantees that the neighborhood
of each vertex is homemorphic to a $4-$ball. This ensures
that any p-simplex in the triangulation is uniquely specified in
terms of its vertices. This differs from
recent work by Laiho \textit{et al.} which utilizes an ensemble of degenerate triangulations
and a different measure term \cite{laihoEvidenceAsymptoticSafety2011,laihoLatticeQuantumGravity2017,daiNewtonianBindingLattice2021,basslerSitterInstantonEuclidean}. Our work is also complementary to that 
of Ambjorn \textit{et al.} \cite{ambjornEuclidian4dQuantum2013} who employ the same class of triangulations but a different
measure term. 

The goal of our work has been to provide a detailed picture of the phase diagram of the model
and the location of possible phase transitions by simulating the model over a fine grid in
the two dimensional parameter space for three lattice volumes ranging 
up to $N_4=32,000$ 4-simplices. We find evidence for a single critical line
separating a crumpled from a branched polymer phase consistent with all
earlier studies of similar models. 
In addition to certain bulk observables we 
have focused
our attention on the Hausdorff and spectral dimensions along this critical line and are able
to compute these both along and transverse to this critical line in some detail.

\section{The lattice model}

\begin{figure*}[!htb]
	\centering
	\hspace{20pt}
\subfloat[\label{fig_chiN0}]{%
	\centering
  \includegraphics[width=.45\textwidth]{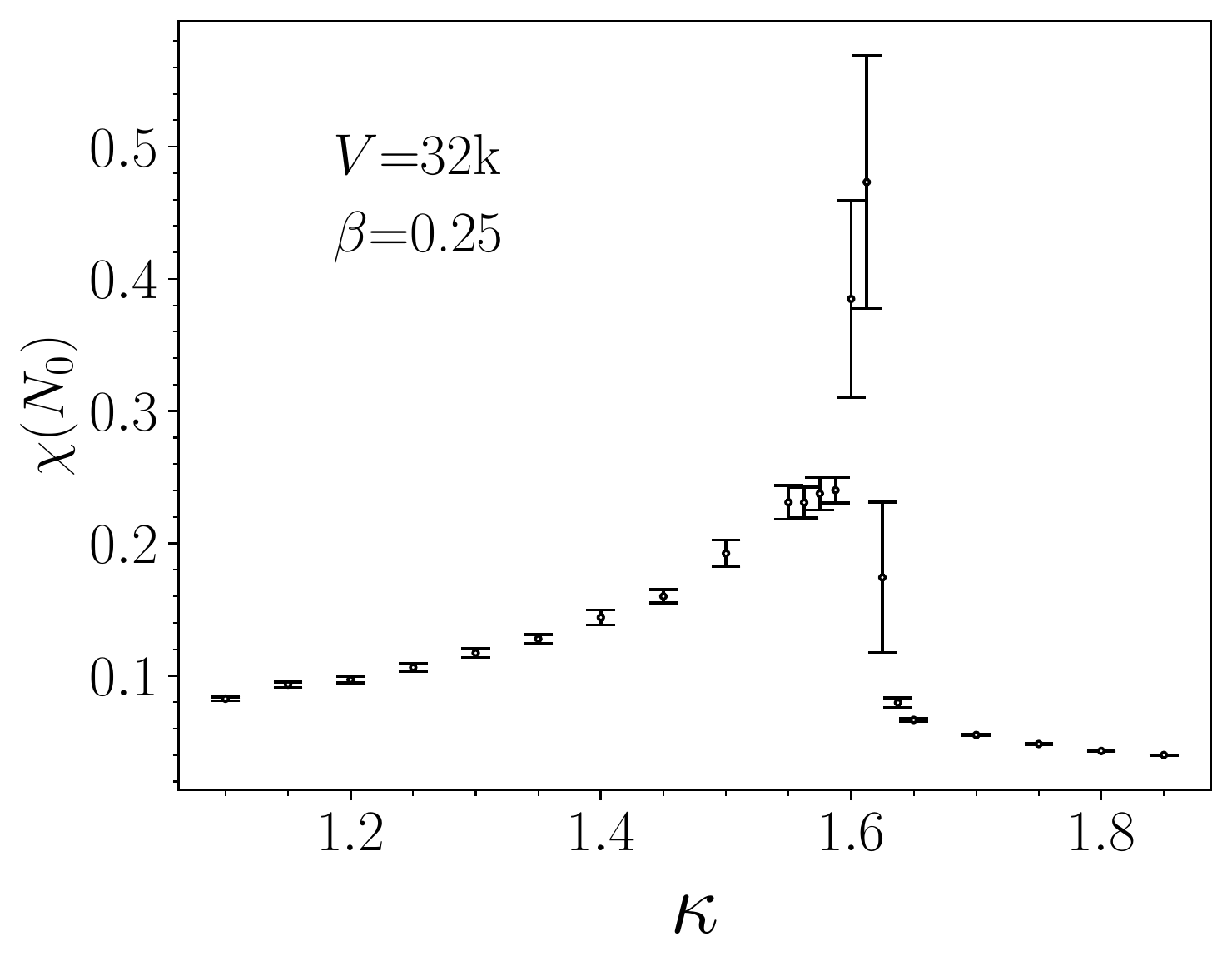}%
}\hfill
\subfloat[\label{fig_chilogq}]{%
	\centering
  \includegraphics[width=.45\textwidth]{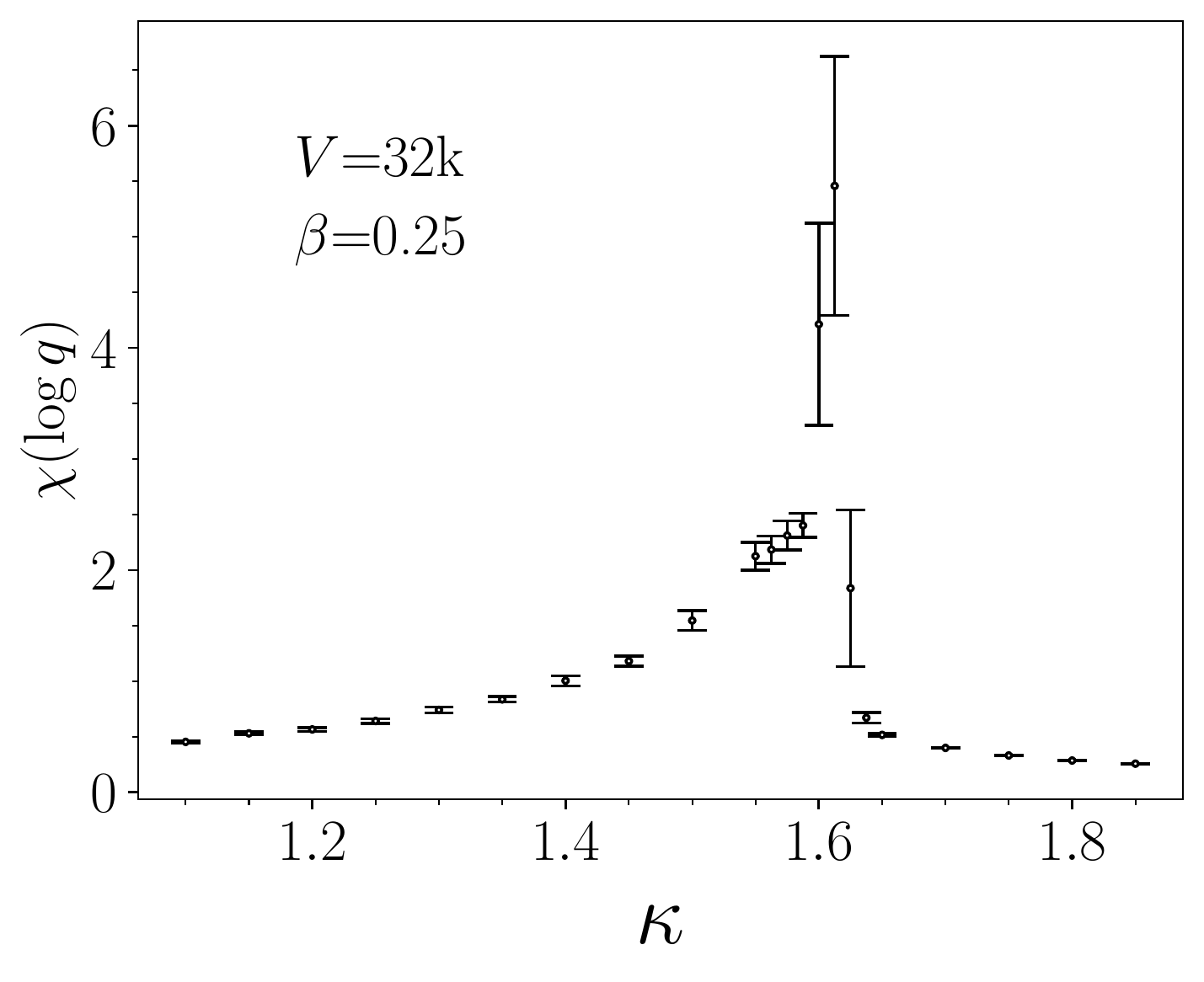}%
}

\hspace{20pt}
\caption{Susceptibility plots (a) $\chi_{N_0}$ and (b) $\chi_{\log{q}}$ for $V=32K$ are shown. The peaked structure near $\kappa \sim 1.6$ indicates a phase transition.}
\label{fig:sus}
\end{figure*}
\begin{figure*}[!htb]
        \centering
        \begin{subfigure}[b]{0.45\textwidth}
            \centering
            \includegraphics[width=\textwidth]{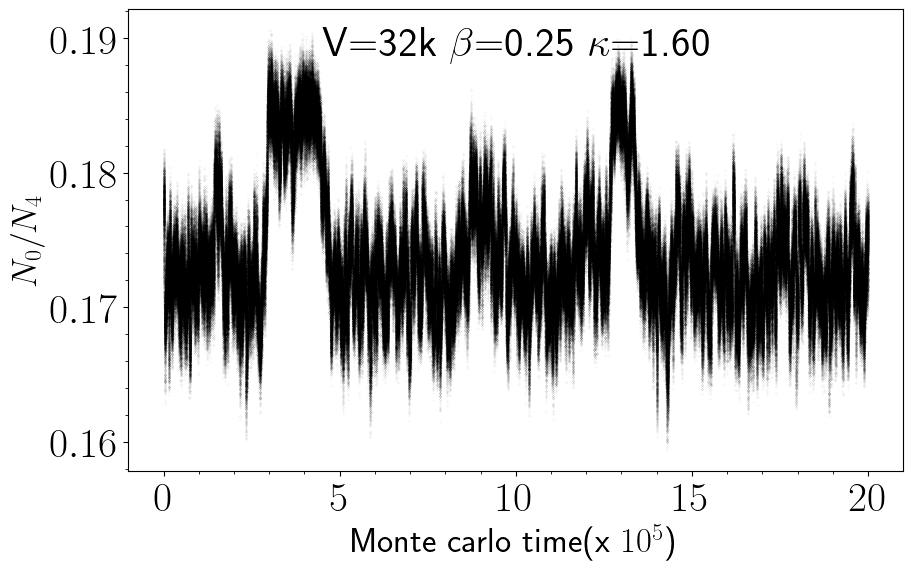}   
        \end{subfigure}
        \hfill
        \begin{subfigure}[b]{0.45\textwidth}  
            \centering 
            \includegraphics[width=\textwidth]{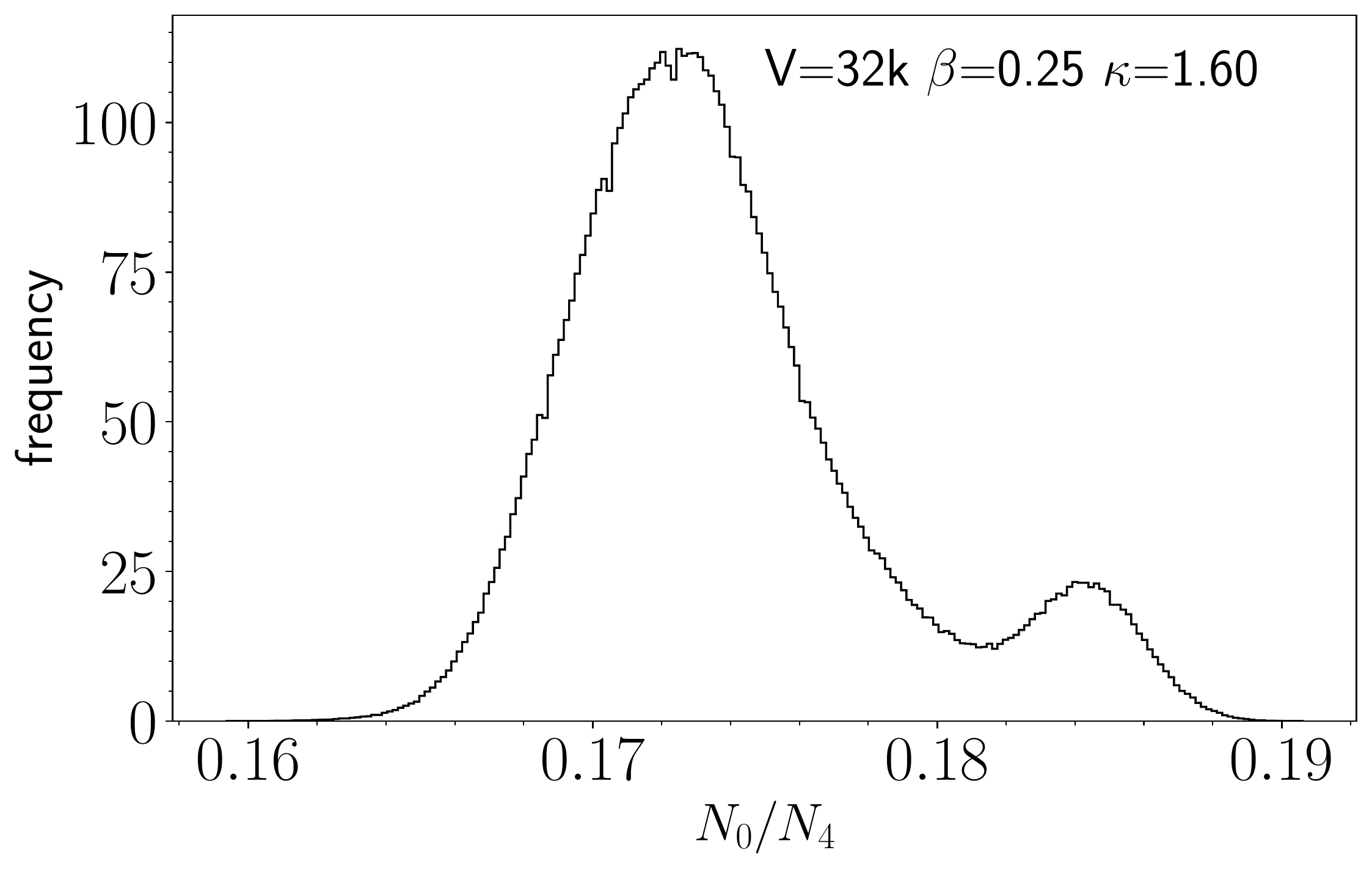}  
        \end{subfigure}
        \caption
        {\small First order nature of the transition at $\beta=0.25$ can be see from the (a) Monte Carlo time series for $N_0/N_4$ and (b) double peak structure of the probability density of $N_0/N_4$.} 
        \label{fig:mc}
    \end{figure*}
The partition function of the model for the pure gravity takes the form
\begin{equation}
    Z=\sum_T \rho(T) e^{-S}\label{partition}
\end{equation}
where the discrete action $S$ is given by
\begin{equation}
    S=-\kappa N_0+\lambda N_4+\gamma(N_4-V)^2,
\end{equation}
where the sum runs over all abstract triangulations
with fixed (here spherical) topology~\footnote{Numerical evidence has been presented in previous studies that the number of possible 4d triangulations of fixed spherical topology is exponentially bounded \cite{catterall1996baby,ambjorn1994exponential} and
hence can be controlled by a bare cosmological constant term.}
The first two terms in the action depending on the number of vertices $N_0$
and the number of four simplices $N_4$ arise from using Regge calculus \cite{regge1961general,thorne2000gravitation} to
discretize the continuum Einstein-Hilbert action with $\kappa$ playing the role
of the bare Newton constant and $\lambda$ a bare cosmological constant. 
The third term plays an auxiliary role in effectively fixing the target volume to $V$ by
tuning $\lambda$ while still allowing for 
small fluctuations $\delta V\sim \frac{1}{\sqrt{\gamma}}$.

The central assumption in this approach to quantum gravity
is that the sum over triangulations
reproduces, in some appropriate continuum limit, the ill-defined continuum path
integral over metrics modulo diffeomorphisms. In two dimensions this prescription is known to reproduce known results
for 2d gravity from Liouville theory and matrix models \cite{boulatov1986analytical,kazakov1986ising} but in higher
dimensions it is merely a plausible ansatz. 

The assumption
of most recent works
is that an additional measure term $\rho(T)$, which
depends on local properties of the triangulation, is needed to ensure this correspondence with continuum gravity remains true \cite{ambjornEuclidian4dQuantum2013,laihoLatticeQuantumGravity2017}.  Here we employ a new form
\begin{equation}
    \rho(T)=\prod_{i=1}^{N_0}q_i^\beta,
\end{equation}
where $q_i$ denotes the number of simplices sharing vertex $i$ and $\beta$ is
a new parameter. This is similar to the local measure term used in previous studies \cite{laihoLatticeQuantumGravity2017,ambjornEuclidian4dQuantum2013}. It is conjectured
that tuning the coupling to such an operator is necessary to restore continuum
symmetries and approach a fixed point where a continuum limit describing quantum
gravity can be taken \cite{ambjornEuclidian4dQuantum2013}. 

Our work is focused on examining the phase structure of the model
in the $(\kappa,\beta)$ with the goal of searching for critical behavior and locating
a region where such a continuum limit can be taken.

\section{Phase Structure}
We employ a Monte Carlo algorithm to sample the sum over random
triangulations \cite{catterall1995simulations}. Five elementary 
local moves (``Pachner moves") which, iterated appropriately
are known to be sufficient to
reach any part of the triangulation space. 

Two of the simplest observables that can be used to locate the transition are
the node and measure susceptibilities which are defined by
\begin{align}
    \chi_{N_0}&=\frac{1}{V}\left(\left\langle N_0^2\right\rangle-\left\langle N_0\right\rangle^2\right)\\
    \chi_{\ln{q}}&=\frac{1}{V}\left(\left\langle Q^2\right\rangle-\left\langle
    Q\right\rangle^2\right)
\end{align}
with $Q=\frac{1}{N_0}\sum_{i=1}^{N_0}\ln{q_i}$.
In fig~\ref{fig:sus} we show these as a function of
$\kappa$ at $\beta=0.25$ for a lattice of (average) volume $N_4=32,000$. The peak in both
quantities indicates the presence of a phase transition.

In fig.~\ref{fig:mc} we show the Monte Carlo time evolution
of the vertex number $N_0$ and its associated probability distribution for a $V=32,000$ simplex simulation close to
the critical line at $\beta=0.25$. The tunneling behavior
in the Monte Carlo time series together with the double peak structure in 
the probability distribution for the number of vertices $P(N_0)$ constitute strong
evidence that the transition is first order in this region. This precludes
a continuum limit and indeed the observation of a similar structure
at $\beta=0$ was the original motivation
for introducing a measure term.
\begin{figure}
    \centering
    \includegraphics[width=0.48\textwidth]{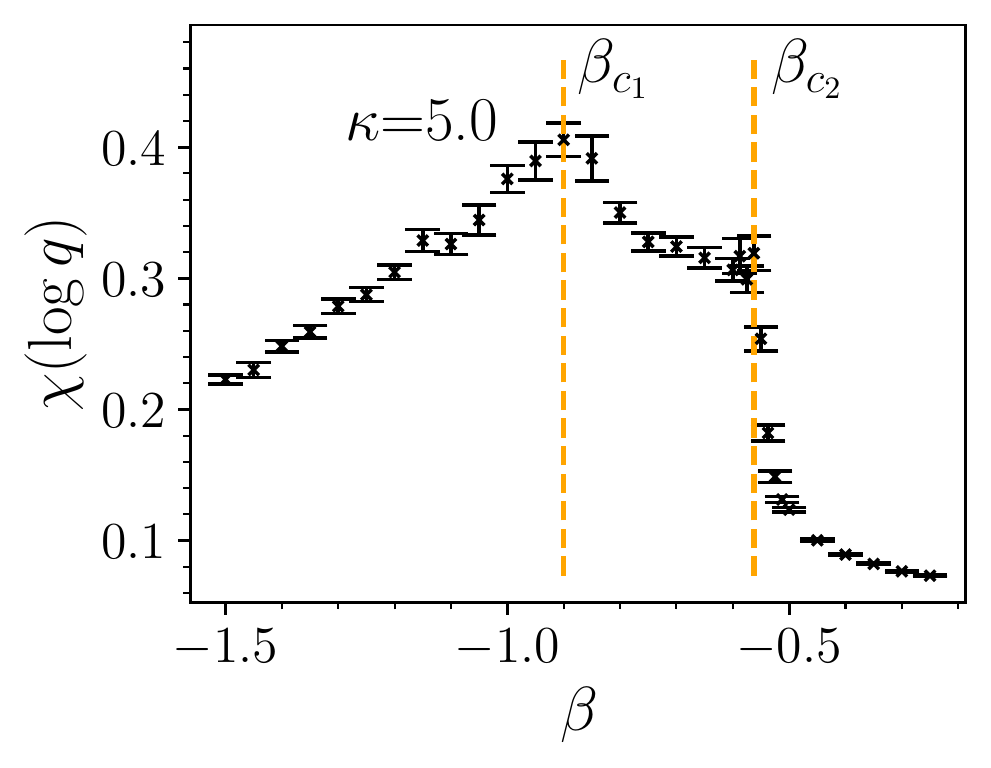}
    \hspace{10pt}
    \caption{At large $\kappa$ two distinctive peaks are observed in the susceptibility plots. Position of the critical points are shown with the vertical lines.}
    \label{broadpeak}
\end{figure}

\begin{figure}
	\centering 
	\includegraphics[width=0.45\textwidth]{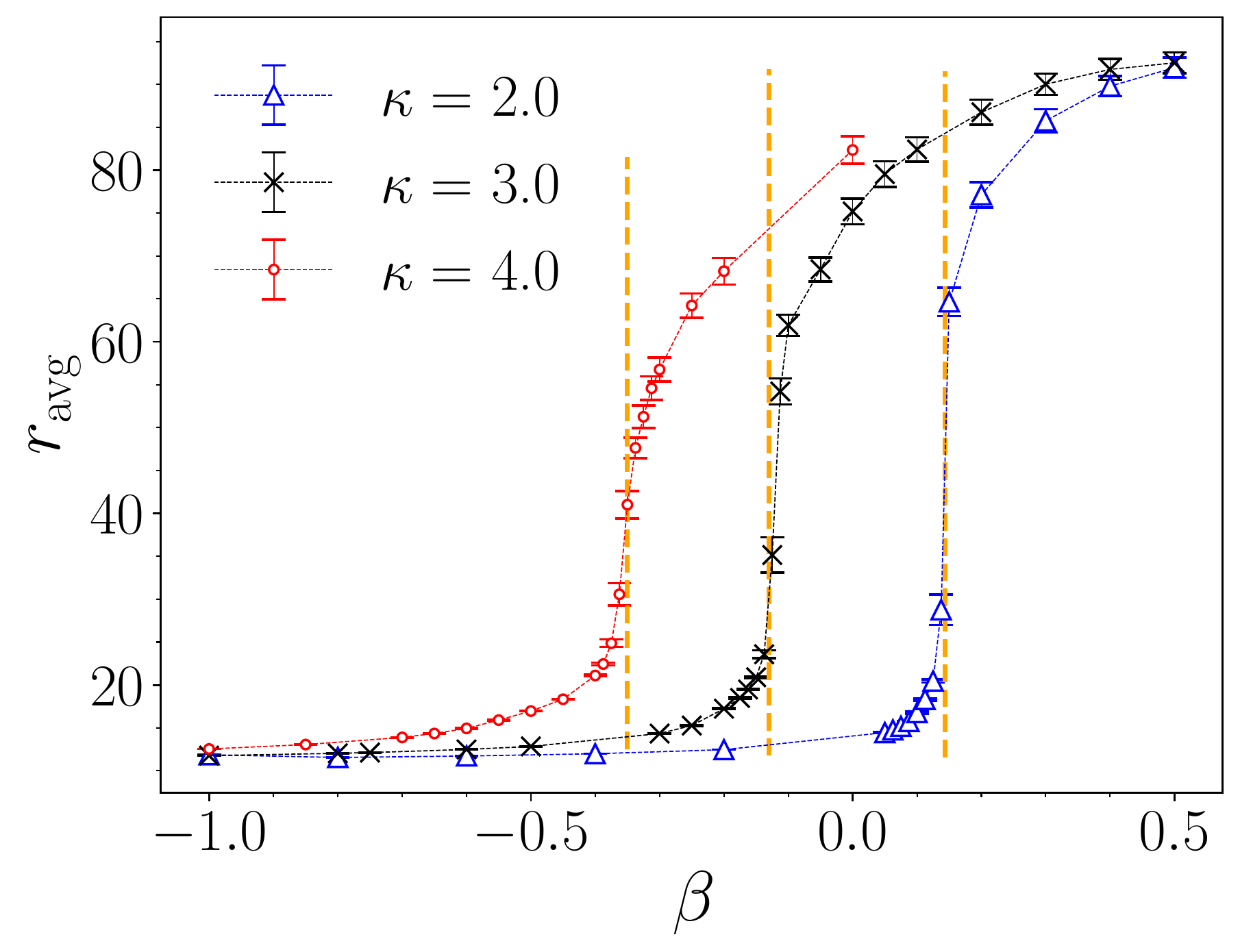}   
	\caption{$\beta$ dependence of the average radius $r_{\rm avg}$ at different fixed values of $\kappa$, at a target volume $V=32$K. Vertical lines denote the position of the critical points $\hat{\beta}_c$ at different Gravitational constants $\kappa$.}
	\label{fig_ravg}
\end{figure}

\begin{table}[!htb]
	\centering
	
	\renewcommand{\arraystretch}{1.5}{
		\begin{tabular}{
				|>{\centering\arraybackslash}p{2.5cm}||
				>{\centering\arraybackslash}p{2.5cm}|
				>{\centering\arraybackslash}p{2.5cm}|
			} 
			\hline
			$\bm{\beta}$ & $\bm{\kappa_c}$  &  $\bm{\hat{\kappa}_c} $\\ \hline
			1.00 & -0.89(1)  & -0.894(6) \\ \hline
			0.50 & 0.75(1) &  0.756(6)\\ \hline 
			0.25 & 1.61(2)  & 1.606(6) \\ \hline \hline
			$\bm \kappa$ & $\bm \beta_c$ & $\bm{\hat{\beta}_c}$\\ \hline
			2.0 & 0.14(1) & 0.144(6)    \\ \hline
			2.5 & 0.00(1) & 0.006(6) \\ \hline
			3.0 & -0.13(1) & -0.13(1)     \\ \hline
			3.5 & -0.25(1) & -0.244(6) \\ \hline
			4.0 & -0.36(1) & -0.35(1)    \\ \hline
			4.5 & -0.46(1)  & -0.46(1)  \\ \hline
			5.0 & -0.56(1)  & -0.56(1)\\ \hline
		\end{tabular}
	}
	\caption{Pseudo-transition point $\kappa_{c}$ ($\beta_c$) obtained from fixed $\beta$ ($\kappa$) scan of the susceptibilities at target volume $V=32$ k vs corresponding
		estimates of the critical point $\hat{\kappa}_c$ ($\hat{\beta}_c$) determined
		from the average radius $r_{\rm avg}$. } \label{tab_TP} 
\end{table}

As we increase $\kappa$ we observe that 
the latent heat of the transition, as measured by the separation in
the two peaks in the probability distribution $P(N_0)$ decreases and the structure
of the susceptibility plots changes. If one fixes $\kappa$
one observes a broad peak centered at $\beta_{c1}$ followed by a much narrower peak at $\beta_{c2}$ with $\beta_{c2}> \beta_{c1}$, Fig.~\ref{broadpeak}. 
For $\beta>\beta_{c2}$ the system is clearly in the branched polymer phase while for $\beta<\beta_{c1}$ the system is clearly in the crumpled phase. The separation $\Delta \beta$ between the two critical points narrows down as the volume is increased. In our work we have used $\beta_{c2}$ as our best estimate for the true critical point $\beta_c$.

To complement this determination of the critical point we have also studied the mean radius of the discrete geometry.
This is defined by
\begin{align}
    r_{\mathrm{avg}}=\frac{1}{N_4}  \left\langle  \sum_r r \,N_{3}(r)   \right\rangle_{T},
\end{align}
where $N_3(r)$ is the
number of four-simplices at geodesic distance $r$ 
measured on the dual lattice from some randomly chosen origin.
In fig.~\ref{fig_ravg} we show a plot of the mean radius $r_{\rm avg}$  vs $\beta$ for
several values of $\kappa$.
The phase transition visible in the susceptibilities
is clearly also seen in $r_{\rm avg}$.  To find the critical coupling, we computed a numerical derivative of the radius as a function of
$\beta$ and identified the critical point $\hat{\beta}_{c}$ as the point where this derivative is maximal. A list of transition points derived from this observable are listed in the second column of the table \ref{tab_TP} and shown to agree very well with the value $\beta_{c}$ determined from the $\chi(N_0)$ and $\chi_{\log{q}}$ susceptibilities. Notice for small $\kappa$ we have fixed $\beta$ and
scanned the transition in $\kappa$ while for large $\kappa$ we have fixed $\kappa$ and done a scan in $\beta$ values\footnote{This was motivated by the schematic phase diagram known from the earlier studies \cite{laihoLatticeQuantumGravity2017}, where the transition line shows trend to asymptote to a negative $\beta_c$ value at large $\kappa_c$. It's true that there is no guarantee that the same trend will be followed in our analysis with the new measure term.}.

\section{Hausdorff dimension}
\begin{figure}[!htb]
\centering
    \begin{subfigure}[b]{0.45\textwidth}
    \centering
    \includegraphics[width=\textwidth]{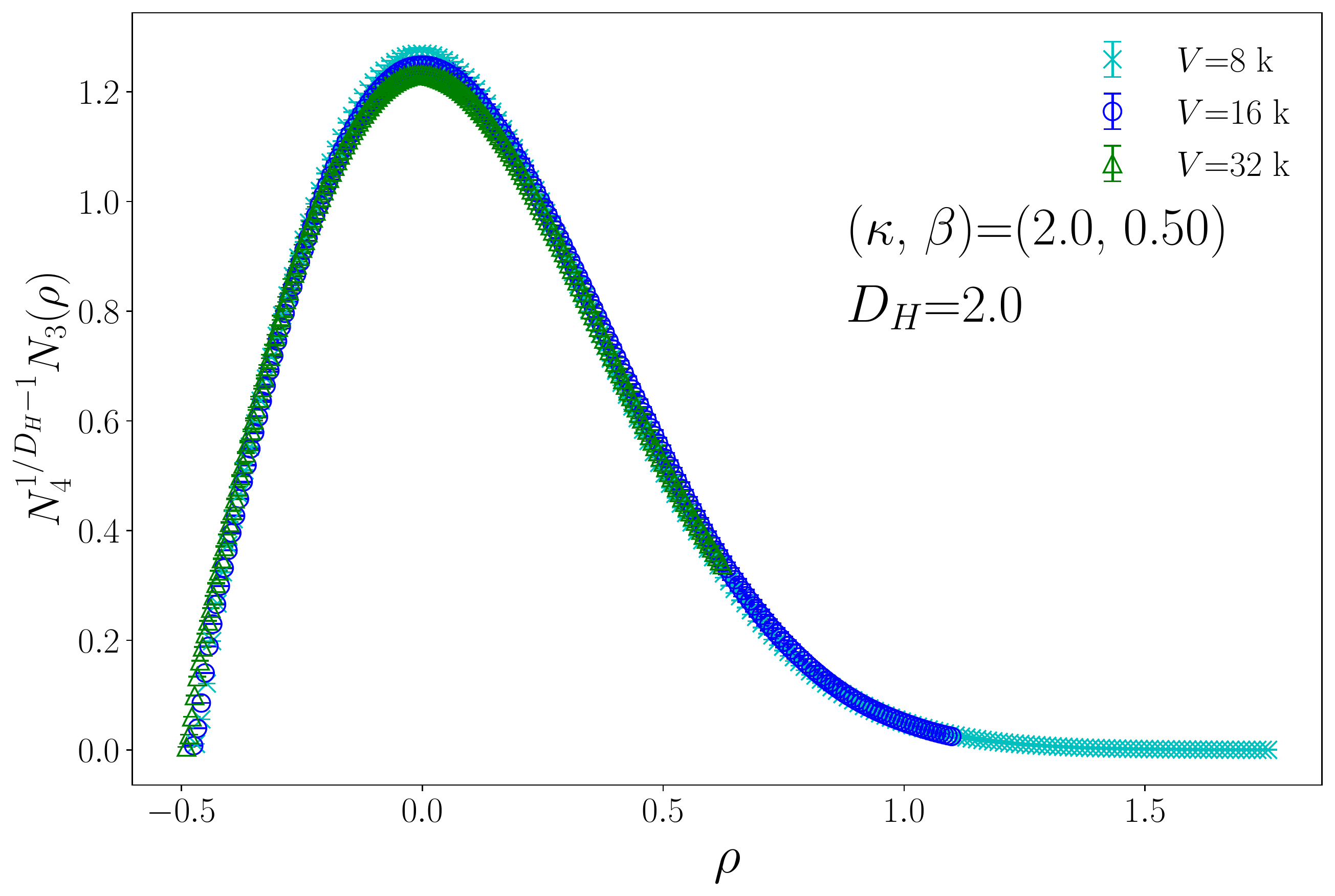}   
    \end{subfigure}
    \caption
    {\small Data collapse of the three volume $N_3(\rho)$ with scaled distance is consistent with $D_H=2.0$ in the BP phase.} 
        \label{fig_BP}
\end{figure}
\begin{figure}[!htbp]
  \includegraphics[width=.35\textheight]{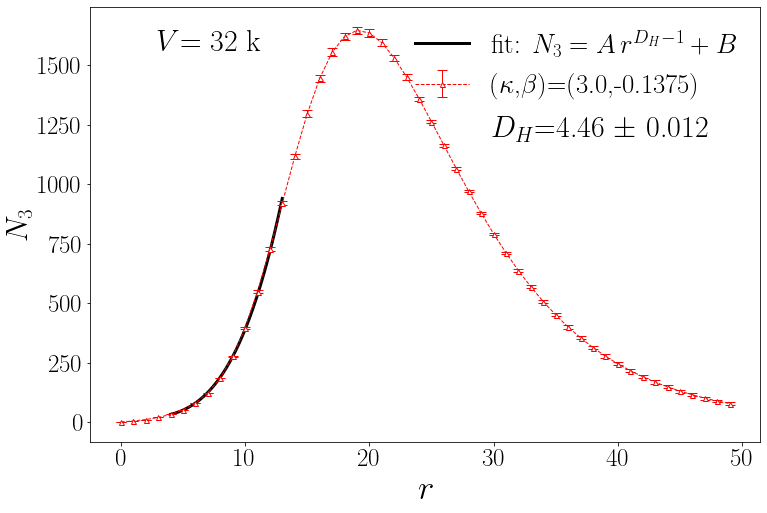}%
 
\caption{Fit of three volume data at small distance.\label{N3fit}}
\end{figure}
\begin{figure}
    \centering
    \includegraphics[width=0.5\textwidth]{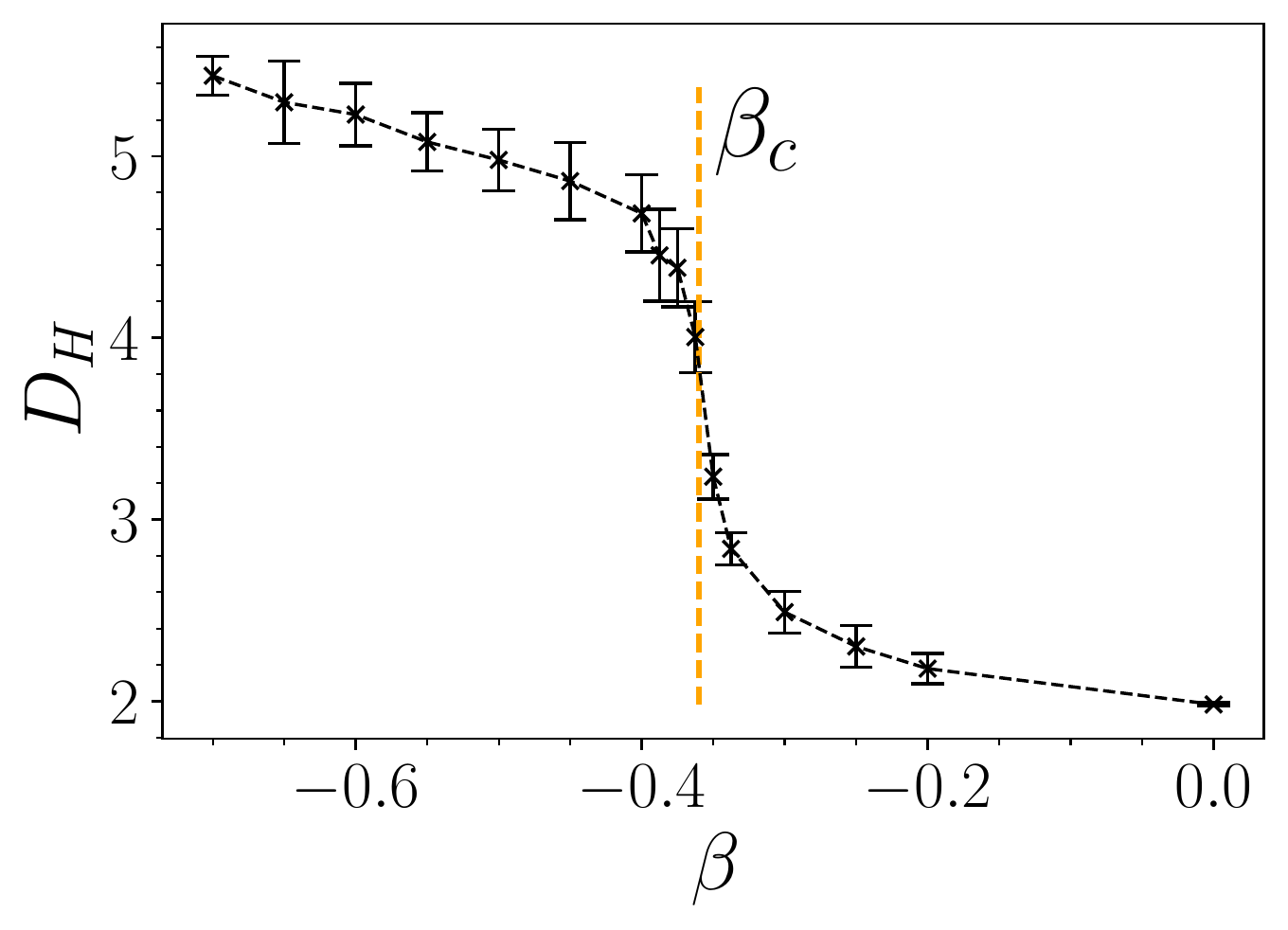}
    \caption{Variation in Hausdorff dimension  $D_H$ with $\beta$
    at $\kappa=4.0$ at $V=32$ K. Position of the critical coupling $\beta_c$ derived from susceptibility is noted with the vertical line. }
    \label{DHscan}
\end{figure}

To compute the Hausdorff dimension $D_H$ we assume that $N_3(r)$ takes the scaling form
\begin{equation}
    N_3 (r)=N_{4}^{1 / D_{H}-1} f\left(r/N_{4}^{1 / D_{H}}\right).
\end{equation}
Fitting to this form shows that
the Hausdorff dimension in the branched polymer phase is consistent with the value of $D_H=2$ (Fig.~\ref{fig_BP})
while in the collapsed phase, the extracted value of $D_H$ from such fits is large which is consistent with the continuum expectation of infinite $D_H$ \cite{ambjorn1995scaling,coumbe2015exploring}.
At small distances, $N_3$ should grow as $\sim r^{D_H-1}$ \cite{ambjorn1995scaling}. In practice we have used this fact rather than data collapse on the scaling form
to extract $D_H$ close to the critical line on our largest lattice by fitting 
\begin{equation}
    N_3=A\,r^{D_H-1} +B.
\end{equation}
Fig~\ref{N3fit} shows such a fit. The results presented are an ensemble average computed from 2000 thermalized configurations. The fit is performed at several fixed $\beta$ and fixed $\kappa$ to observe the variation in the Hausdorff dimension as we moved from the collapsed phase to the branched polymer phase. The value of the Hausdorff dimension is strongly influenced by
the distance from the critical line as can be seen in Fig.~\ref{DHscan} which shows $D_H(\beta)$ at a fixed $\kappa=4.0$. From the rise of the value of DH towards the left, it is evident that as we probe deep into the collapsed phase, we get larger Hausdorff dimensions. Also clearly visible is the
fact that deep in the BP phase on the right of the diagram the value approaches the known value of $D_H=2$.

\begin{figure}[!htb]
    \centering
    \includegraphics[width=.45\textwidth]{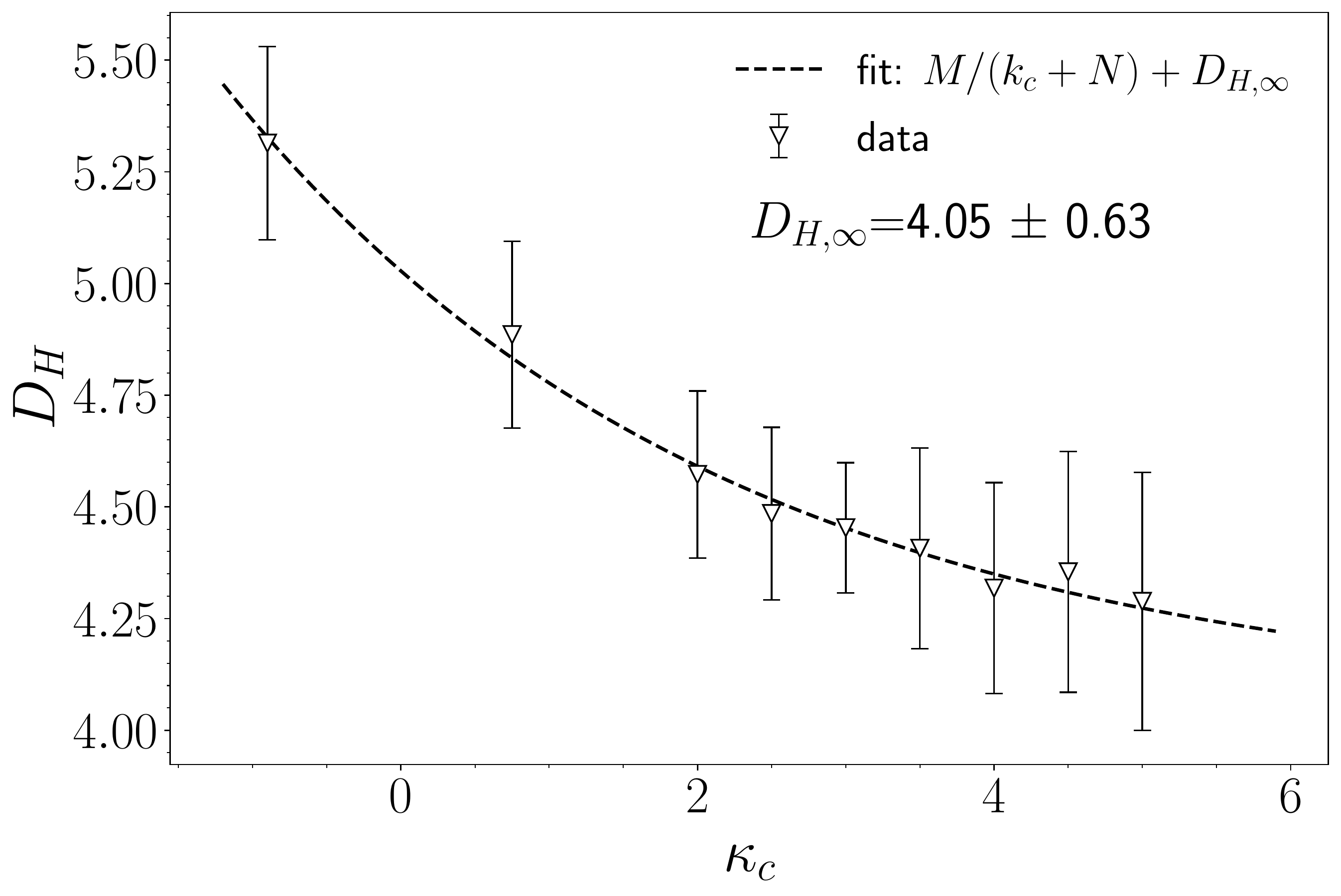}%
    \caption{Fit of extracted Hausdorff dimension $D_H$ as a function of critical coupling along transition line.}
   \label{DHextrap}
\end{figure}
The value of  $D_H$ along the critical line is shown in fig.~\ref{DHextrap} which also
includes a fit of the form
\begin{equation}
    D_H=M/(\kappa+B)+D_{H,\infty}.
\end{equation}
Here, $M$ and $B$ are fit parameters and $D_H \to D_{H,\infty}$ as $\kappa\to\infty$. We find $D_{H,\infty}=4.06 \pm 0.64$ which 
is consistent with the emergence of four dimensional de Sitter space in this limit.

\begin{figure}
    \centering
    \includegraphics[width=0.45\textwidth]{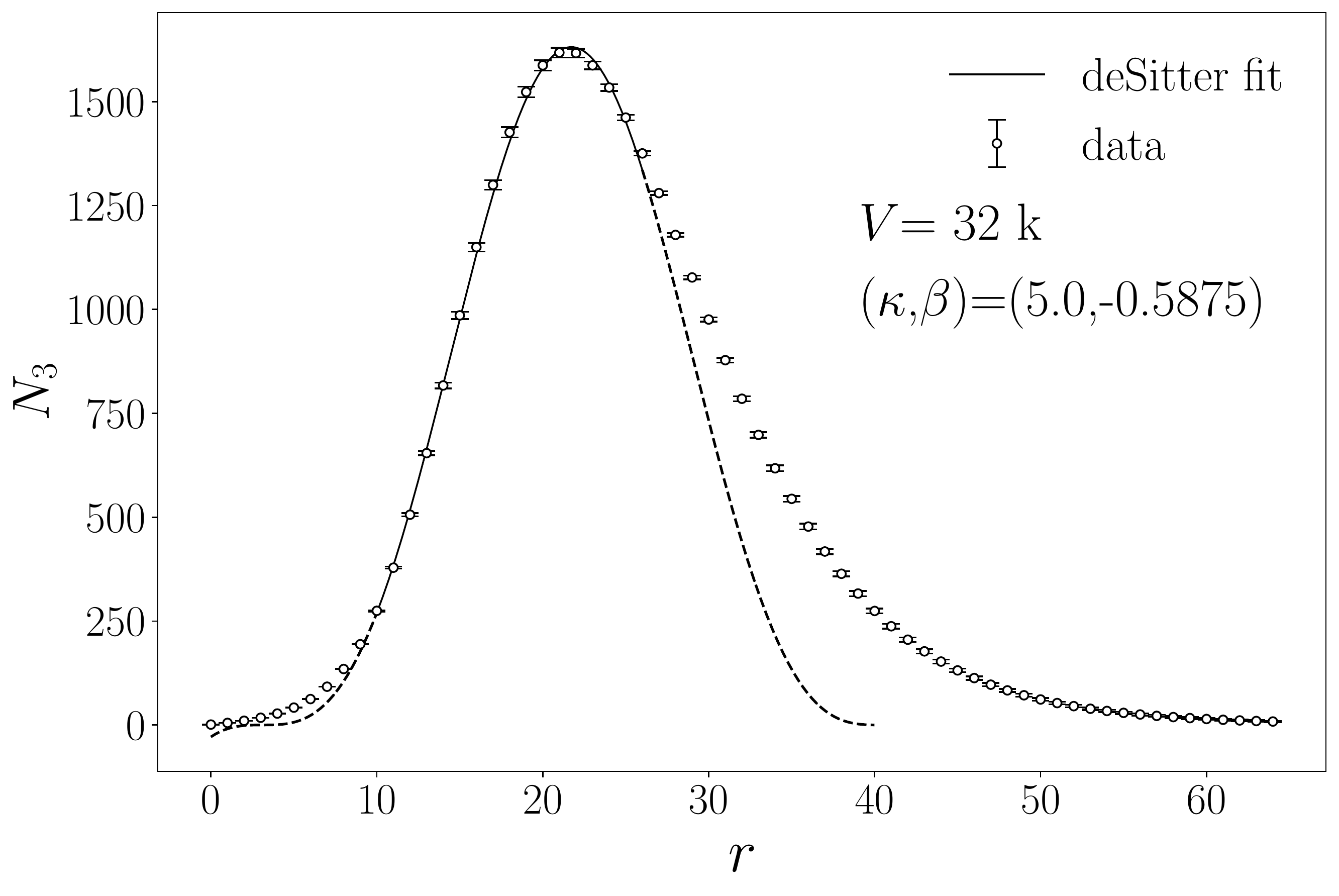}
    \caption{Fit to the de-Sittter solution of the three volume distribution data.}
    \label{deSitterfit}
\end{figure}

Encouraged by
this we have compared our three-volume distribution  near the
critical point at large $\kappa$ with the (Euclidean) de-Sitter solution  \footnote{A homogenous and isotropic universe as described by the FLRW metric}. The
associated three volume profile for the Wick rotated case takes the form of Eqn.~\ref{desitter_3vol} \cite{ambjorn2008nonperturbative,glaser2017cdt,laihoLatticeQuantumGravity2017}.  This is shown in Fig.~\ref{deSitterfit} and indicates that the average geometry at small to intermediate
distances is indeed consistent
with de Sitter as $\kappa$ gets large.
\begin{equation}
N_{3} (r)=\frac{3}{4} N_{4}^{3/4} \Gamma \cos ^{3}\left(\frac{r-b}{s_{0} N_{4}^{1 / 4}} \right) \label{desitter_3vol}
\end{equation}
Here, $s_0$, $\Gamma$ and $b$ are fit parameters. One can think of
$s_0$ as determining a relative lattice spacing for different values of
the $(\kappa,\beta)$. We find good matching of our data to the de-Sitter solution starting from a small distance $r$ up to about five steps beyond the maxima. The long tail of the distribution is likely
a finite size effect \cite{laihoLatticeQuantumGravity2017}.

\section{Spectral Dimension}
\begin{figure}[!htb]
	\centering
	\includegraphics[width=0.45\textwidth]{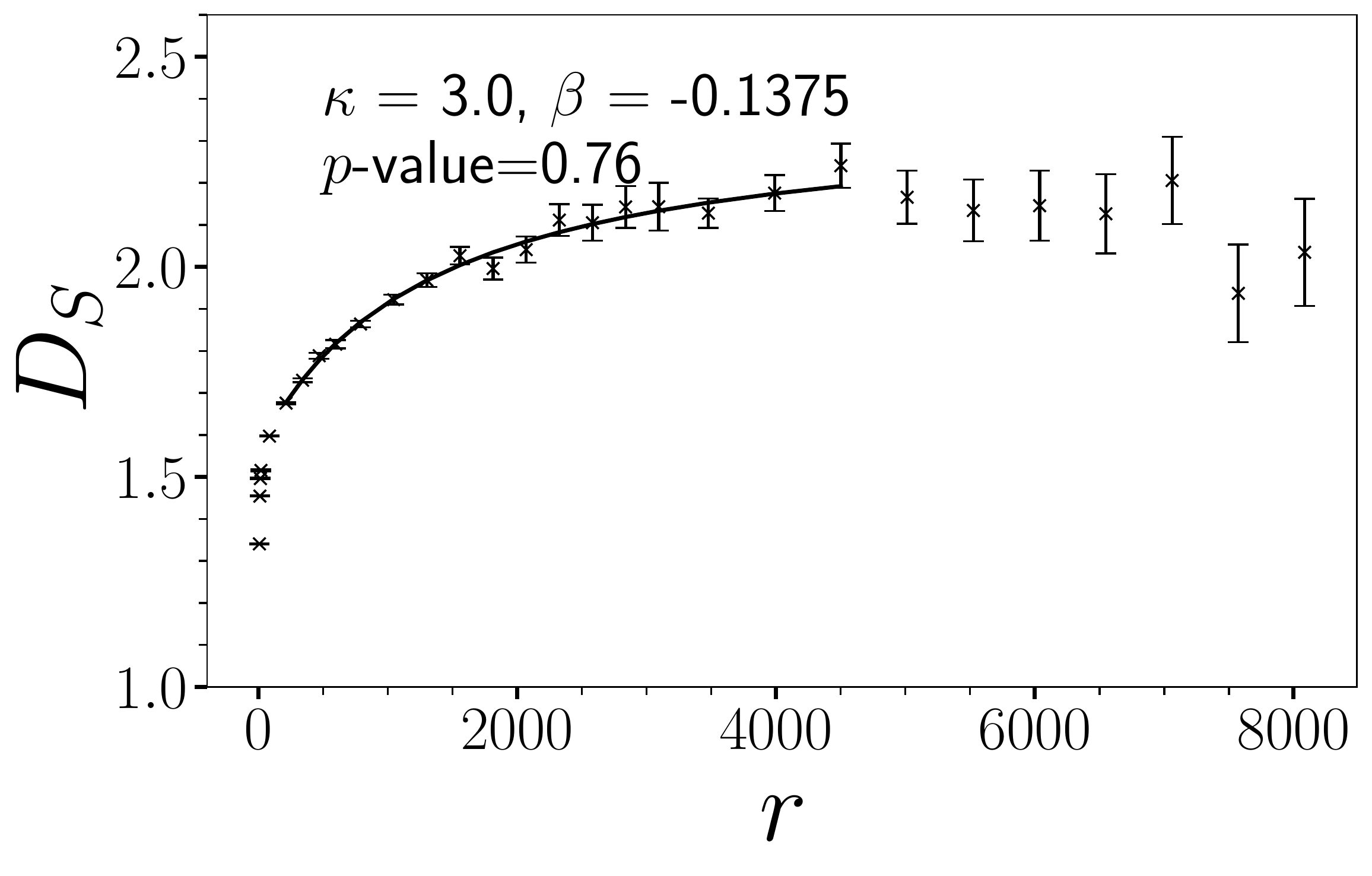}
	\caption{Sample fit of the spectral dimension near the transition line in the phase space.}
	\label{spectral_fit}
\end{figure}

\begin{figure}[!htb]
	\centering
	\includegraphics[width=0.45\textwidth]{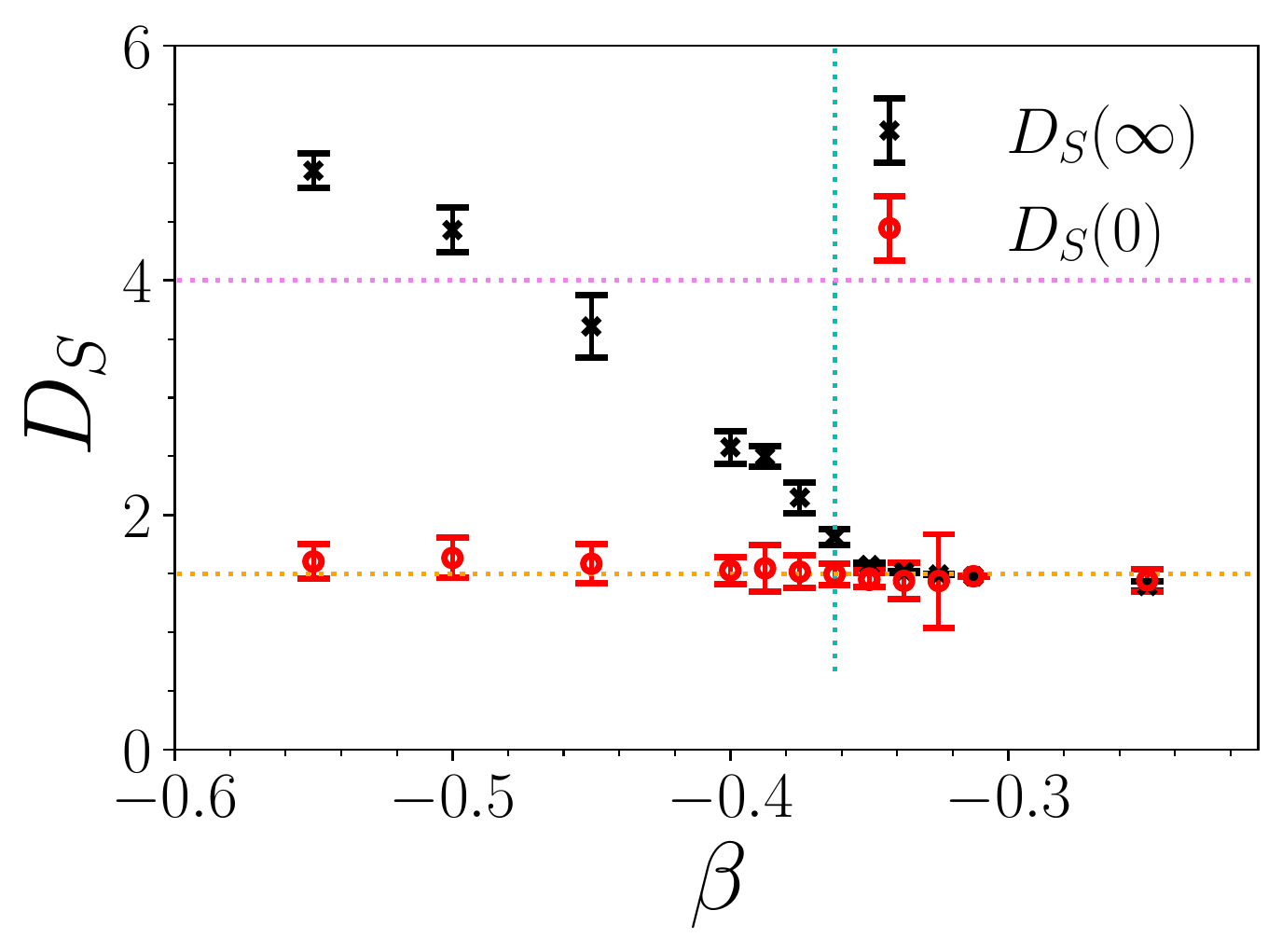}   
	\caption
	{\small UV ($D_S(0)$) and IR ($D_S(\infty)$) spectral dimension across transition at a fixed $\kappa=4.0$. Vertical line denotes the position of the transition point and the two horizontal line denotes the $D_s$ value of 1.5 and 4 for comparison with the data.} 
	\label{fig_DSinf0_across}
\end{figure}

\begin{figure}[!htb]
	\centering 
	\includegraphics[width=0.45\textwidth]{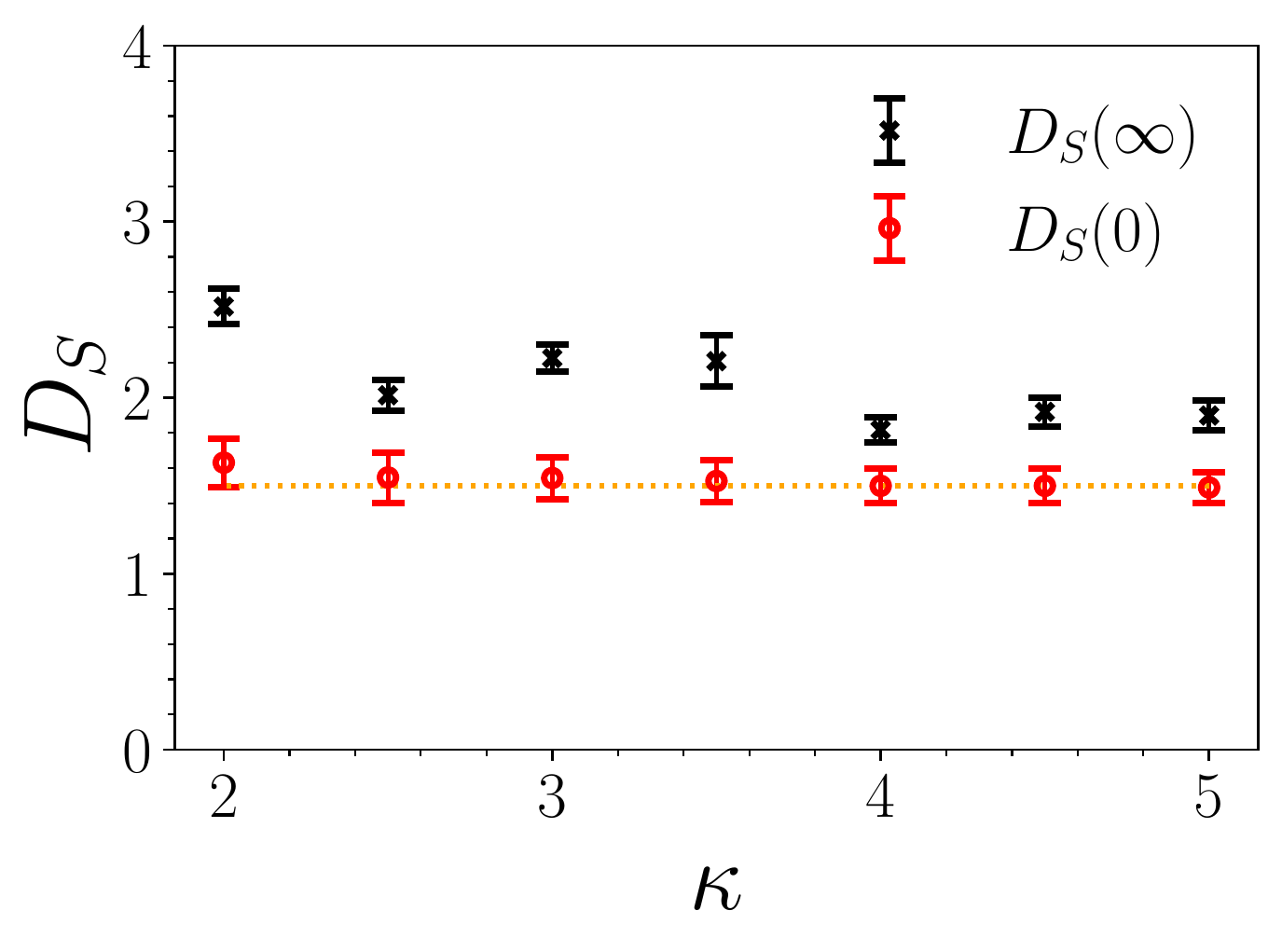}  
	\caption
	{\small UV ($D_S(0)$) and IR ($D_S(\infty)$) spectral dimension along the the transition line. Horizontal line denotes the $D_s$ value of 1.5 for the comparison.} 
	\label{fig_DSinf0_along}
\end{figure}
Another measure of
dimension for a fluctuating geometry is called the ``spectral dimension" $D_S$. 
It can be computed
for a simplicial manifold $\mathcal{M}$ using a random walk process. 
Starting from a randomly selected
simplex the random walk corresponds to successively moving from one simplex
to one of its neighbors
via a randomly selected face. This process is then
iterated a large number of times. To compute the spectral dimension one records the
number of times the walk returns to the starting simplex as a function of the diffusion
time (number of steps of the random walk).
By running several of these walks and averaging over starting
points and over the ensemble of configurations obtained at
some fixed $\beta$ and $\kappa$ 
we can obtain the probability of returning to
the starting simplex $P_r(\sigma)$ after $\sigma$ steps. The spectral
dimension is then defined from the relation:
\begin{equation}
D_{S}(\sigma)=-2 \frac{d \log \left\langle P_{r}(\sigma)\right\rangle}{d \log \sigma}
\end{equation}
The return probability itself is a useful quantity which can be used to find the
relative lattice spacing at different points on the transition line \cite{laihoLatticeQuantumGravity2017}. This is discussed in
more detail in appendix~\ref{appendix1}.

In the branched polymer phase we observe $D_S=4/3$ which is consistent
with theoretical expectations \cite{jonsson1998spectral} while in the crumpled phase it is
large. At the critical point we find $D_S$ is not well fitted by
a constant but instead runs with scale $\sigma$.
In fig.~\ref{spectral_fit} we show a plot of this running spectral dimension for $V=32K$ and $\beta=-0.1375,\,\, \kappa=3.0$. \\

We used 2000 thermalized configurations for the computation of the spectral dimension. Each random walk is performed up to 15000 steps and we choose 32000 randomly chosen sources per configuration. The fit is attempted over different ranges. Due to the finite volume of the lattices, the spectral dimension will increase and reach a maximum before decreasing. However, the number of steps needed to
reach this maximum depends on the effective dimension of the manifold. We attempted to fit our data up to this maximum whenever possible. This amounts to choosing different fit ranges at different regions of the parameter space. The choice of the fit range can be justified by tracking the p-value of the fits.\\

As in previous works \cite{ambjorn2005spectral,laihoLatticeQuantumGravity2017}, we found the following fit function best represents the data
\begin{equation}
    D_S (\sigma)=a+\frac{b}{c+\sigma}.
\end{equation}
The fit function yields estimates for the spectral dimension at small distances $D_S(0)$ and also at large distances $D_S(\infty)$. A single elimination jackknife procedure is used to compute the error-bar, and the fit is performed for different fit ranges. Systematic errors due to the choice of the fit range are added in quadrature with the statistical error of the best fit used to compute the overall error. We use the metric `p-value' to select reasonable fit ranges for the data. Fig.~\ref{fig_DSinf0_across} shows the variation of $D_S(0)$ and $D_S(\infty)$  across the transition line from the
crumpled to the branched polymer phase, while Fig.~\ref{fig_DSinf0_along} shows 
the variation of these quantities along the transition line.
Clearly $D_S$ runs to small ($D_S\sim 1.5$) values in the UV which is consistent with the earlier EDT studies \cite{laiho2017recent}, and CDT studies \cite{coumbeEvidenceAsymptoticSafety2015}. In the IR regime the spectral dimension $D_S(\infty)$ is larger with $D_S(\infty)$ varying from $1.82-2.52$. This
scale dependence of the spectral dimension was also seen earlier in CDT \cite{ambjorn2005spectral}, renormalization group approach \cite{lauscher2005fractal}, loop quantum gravity \cite{modesto2009fractal} and in string theory models \cite{hovrava2009spectral}. 
It is not clear from our study whether the UV spectral dimension $D_S(\infty)$ attains larger values for larger $N_4$. Larger volume simulations with combinatorial triangulations must be conducted to resolve the tension in $D_S(\infty)$ with the results obtained from the degenerate combinatorial calculations \cite{laihoEvidenceAsymptoticSafety2011}~\footnote{In this
work we haven't performed a double scaling of this quantity
using both the lattice volume and the 
lattice spacing as suggested by Laiho \textit{et. el.} \cite{laihoLatticeQuantumGravity2017}. Performing such an
extrapolation might be important for extracting
a continuum value for the UV spectral dimension.}

\section{Conclusions}
We have explored the phase diagram of combinatorial Euclidean dynamical
triangulation models of four dimensional quantum gravity. Our model
contains two parameters - a bare gravitational coupling $\kappa$
and a measure parameter $\beta$.
We find evidence for a critical line 
$\kappa_c(\beta)$ dividing a crumpled phase from a branched polymer phase in 
agreement with earlier work \cite{ambjornEuclidian4dQuantum2013,laihoLatticeQuantumGravity2017}. While this line is associated with first order phase
transitions for small $\kappa$ this transition softens with increasing coupling. An intermediate
``crinkled" phase opens up in this regime but we have focused our attention on the
boundary between this region and the branched polymer phase in our analysis since
this is the only place where we have observed consistent scaling that survives the
large volume limit. When we refer to the critical point in our results we always mean
the boundary between the crinkled and branched polymer phases.

The focus of much of our work has been
to compute the Hausdorff and spectral dimensions as we approach this critical
line from the crumpled phase. We find
evidence that the Hausdorff dimension $D_H$ along the critical
line approaches $D_H=4$ as $\kappa$ increases where it is possible to obtain
increasingly good fits to
to classical de Sitter space. The spectral
dimension $D_S(s)$ is observed to run with 
scale $s$ attaining values consistent with $D_S(0)=\frac{3}{2}$ at short distances
for all values
of $\kappa$. These results are consistent with earlier work using degenerate triangulations
and causal dynamical triangulation models and models using different measure terms \cite{ambjornEuclidian4dQuantum2013, laihoLatticeQuantumGravity2017,coumbeEvidenceAsymptoticSafety2015}.
However our measurement of the spectral dimension at long distances
$D_s(\infty)$ barely exceeds $D_s(\infty)\sim 2$. This result is somewhat in tension with
the earlier work. However, 
we show that $D_s(\infty)$ depends strongly on the distance in parameter space from
the critical line  which renders such measurements delicate and may explain this
discrepancy. Large finite volume effects which have been observed in earlier
studies may also make this measurement difficult.

\FloatBarrier

\acknowledgments 
We acknowledge Syracuse University HTC Campus Grid and NSF award ACI-1341006
for the use of the computing resources. S.C was supported by DOE grant DE-SC0009998.

\appendix

\section{Relative lattice spacing \label{appendix1}}
\begin{figure}[!ht]
	\vspace{12pt}
   \centering
   \begin{subfigure}{0.45\textwidth}
       \centering
       \includegraphics[width=\textwidth]{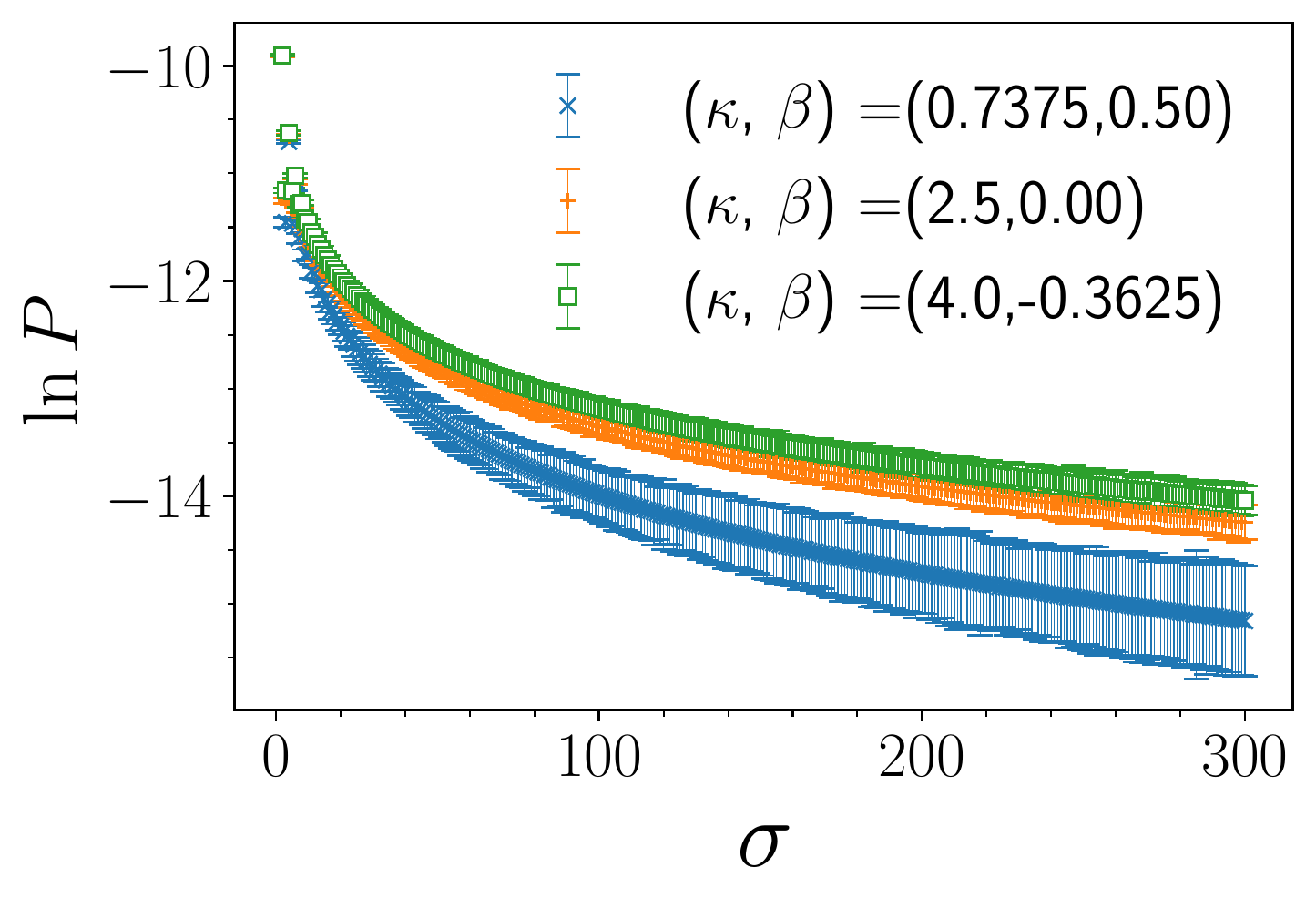}
       \subcaption{\label{fig_ret_prob}}
   \end{subfigure}
   \vfill
   \begin{subfigure}{0.45\textwidth}
       \centering
       \includegraphics[width=\textwidth]{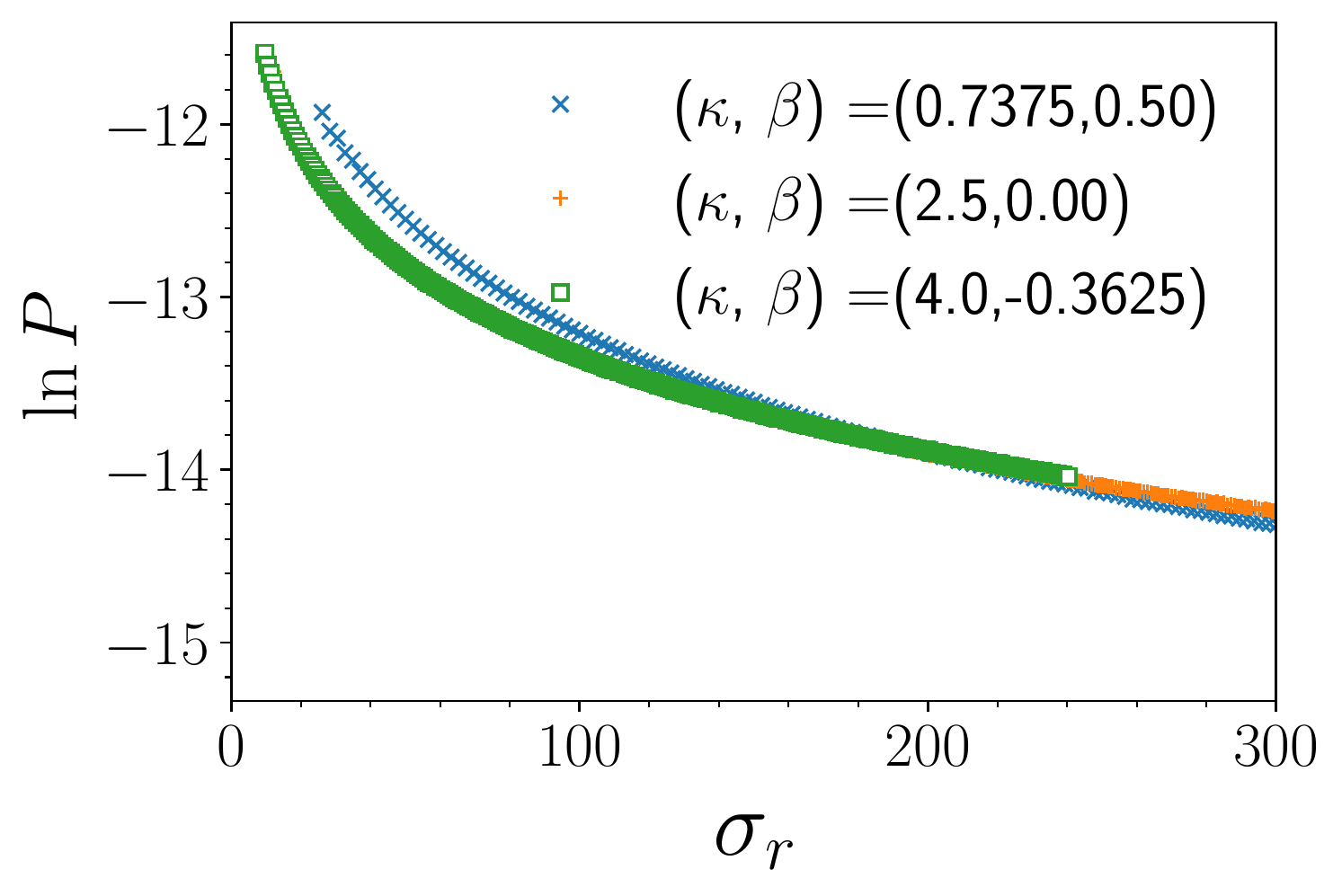}
       \subcaption{\label{fig_rel_a}}
   \end{subfigure}
    \caption{ Return probability at different points in the critical line at lattice volume $V=32$ k (a) with respect to diffusion step ($\sigma$), (b) with respect to rescaled diffusion step ($\sigma_r$). The scaling allows to find relative lattice spacing along the transition line. Associated error-bars are not shown in the right-hand figure to demonstrate the superimposed data from different points in the transition line.  \label{relative} }
\end{figure}

\begin{table}[!ht]
	\centering
	\begin{tabular}{|c| c|c| c|c|c|c|  c|c|c|c|}\hline
		$\kappa_c$ & -0.90& -0.7375 & 1.5625 & 2.0 &2.5 & 3.0 &3.5  & 4.0 &4.5 &5.0 \\ \hline
		$a_r$ & 1.5 & 1.475 & 1.135 & 1.05 &1 &0.935&0.92&0.895&0.87 & 0.86\\ \hline 
	\end{tabular}
	\caption{Relative lattice spacing along the transition line as $\kappa$ is varied.}
	\label{tab_a_r}
\end{table}
In this work, we did not attempt to perform a precise measurement of the renormalized gravitational constant which determines the absolute lattice spacing. Two different methods of finding the gravitational constant in the context of the Euclidean dynamical triangulation can be found in the two recent papers by Laiho \textit{et.el.} \cite{dai2021newtonian,bassler2021sitter}. We have
used the relative lattice spacing as obtained from the return probability in our work.
In fig.~\ref{fig_ret_prob} we show the return probability at several different
points along the critical curve for $V=32$ K and in fig.~\ref{fig_rel_a} and show how
these curves can be collapsed onto a single curve by rescaling the step size
$\sigma$. Rescaling of the step $\sigma=\sigma_r a_r^2$ can be interpreted as yielding a relative lattice spacing $a_r$ as $\kappa$ varies along the critical curve. Values of the relative lattice constant $a_r$ are noted in the Table~\ref{tab_a_r} and are consistent with the previous work by Laiho \textit{et. el.}. Namely they reveal that as $\kappa$ approaches infinity the
corresponding lattices get finer \cite{laihoLatticeQuantumGravity2017}. Hence, for a fixed target volume $V$, the physical volume is smaller at larger $\kappa_c$ and it is likely that the results obtained would suffer greater finite size effect in that region.

\FloatBarrier

\bibliography{dt}

\end{document}